\documentstyle[aps,epsf]{revtex}
\begin{document}
\draft
\tighten
\title{Dominance of Pion-exchange in R-parity Violating Supersymmetry
       Contributions to Neutrinoless Double Beta Decay}
\author{
Amand Faessler$^1$,  Sergey Kovalenko$^{1,2}$,
Fedor \v Simkovic$^{1,2,3}$
and Joerg Schwieger$^1$}
\address{
1. Institut f\"ur Theoretische Physik der Universit\"at T\"ubingen,\\
   Auf der Morgenstelle 14, D-72076 T\"ubingen, Germany\\
2. Joint Institute for Nuclear Research, 141980 Dubna, Russia \\
3. Department of Nuclear Physics, Comenius University, \\
   Mlynsk\'a dolina F1, 84215 Bratislava, Slovakia}
\date{\today}
\maketitle
\begin{abstract}
We present a new contribution of the R-parity violating 
($R_p \hspace{-1em}/\;\:$) supersymmetry (SUSY)
to neutrinoless double beta decay ($0\nu\beta\beta$)
via the pion exchange between decaying neutrons.

The pion coupling to the final state electrons is induced by 
the $R_p \hspace{-1em}/\;\:$ SUSY interactions.
We have found this pion-exchange mechanism to dominate
over the conventional two-nucleon one.
The latter corresponds to direct interaction between
quarks from two decaying neutrons without any light hadronic mediator
like $\pi$-meson.

The constraints on the certain $R_p \hspace{-1em}/\;\:$
 SUSY parameters are extracted from
the current experimental $0\nu\beta\beta$-decay half-life limit.
These constraints
are significantly stronger than those previously known or expected
from the ongoing accelerator experiments.
\end{abstract}

Neutrinoless double beta decay (${0\nu\beta\beta}$)
has long been recognized as
a sensitive probe of the new physics beyond the standard model (SM)
(see \cite{hax84}-~\cite{doi85}).
Various mechanisms of  ${0\nu\beta\beta}$ decay were proposed and studied
in the last two decades.
The conventional mechanism is based on the exchange of
a massive Majorana neutrino between the two decaying neutrons.
A new mechanism was found within supersymmetric (SUSY)
models with R-parity violation ($R_p \hspace{-1em}/\;\:$)
in  \cite{Mohapatra}.
($R_p = (-1)^{3B+L+ 2S}$ where $S, \ B$ and $L$ are spin,
baryon and lepton numbers.)
It was later studied in more details in  \cite{Vergados}.
A complete analysis of this mechanism within the minimal supersymmetric
standard model (MSSM) was carried out in  \cite{HKK}.

\begin{figure}[htb]
\vspace{1.cm}
\hspace{1.5cm}
\mbox{\epsfxsize=13. cm\epsffile{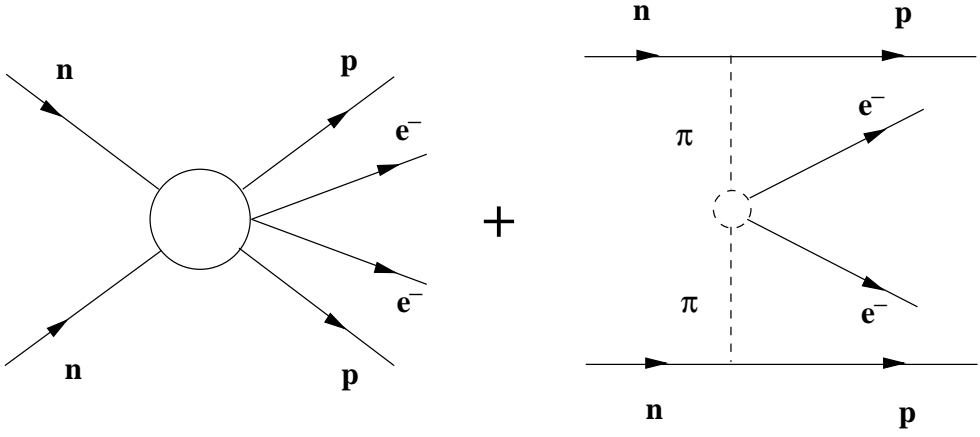}}
\vspace{0.5cm}
\caption{
(a) Two-nucleon mode ${\cal R}_{{0\nu\beta\beta}}^{2N}$ and
(b) $\pi$-exchange ${\cal R}_{{0\nu\beta\beta}}^{\pi N}$
   contributions to ${0\nu\beta\beta}$-decay matrix element
${\cal R}_{{0\nu\beta\beta}}
= {\cal R}_{{0\nu\beta\beta}}^{2N} + {\cal R}_{{0\nu\beta\beta}}^{\pi N}$.}
\label{fig1}
\end{figure}

The nuclear ${0\nu\beta\beta}$-decay
is triggered by the ${0\nu\beta\beta}$ quark transition
$d + d\rightarrow u + u + 2 e^-$ which is induced by certain
fundamental interactions.
It was a common practice to put the initial d-quarks separately
inside the two initial neutrons of a ${0\nu\beta\beta}$-decaying nucleus.
This is the so called two-nucleon mode of the ${0\nu\beta\beta}$-decay
[see Fig.\ \ref{fig1}(a)].
If the above ${0\nu\beta\beta}$ quark transition proceeds at short distances,
as in the case of $R_p \hspace{-1em}/\;\:$ SUSY interactions,
then the basic nucleon transition  amplitude
$n + n\rightarrow p + p + 2e^-$ is strongly suppressed
for relative distances larger than the mean nucleon
radius.

In this letter we propose a new pion-exchange SUSY mechanism which is
based on the double-pion exchange between the decaying neutrons
[Fig.\ \ref{fig1}(b)].
At the quark level this mechanism implies the same short-distance
$R_p \hspace{-1em}/\;\:$
 MSSM interactions as in  \cite{HKK}. However, it essentially differs
from the previous consideration of the SUSY contribution
to the ${0\nu\beta\beta}$-decay at the stage of the hadronization.
We assume that the $R_p \hspace{-1em}/\;\:$ MSSM quark interactions induce
$\pi\pi\rightarrow 2e$ transition at the middle point of the diagram
in Fig.\ \ref{fig1}(b).
The importance of the pion-exchange currents in ${0\nu\beta\beta}$-decay
was first been pointed out by B. Pontecorvo ~\cite{Bruno}.
Latter, this idea was quantitatively realized in
 ~\cite{Vergados1}-~\cite{Fedia} for the case of the
heavy Majorana neutrino exchange. It was shown that
the pion-exchange contribution can not be neglected
in this case.
We will show that
in the case of the $R_p \hspace{-1em}/\;\:$ MSSM induced quark transition
the pion-exchange contribution absolutely dominates over
the conventional two-nucleon mode.

The $R_p$-violating part of the superpotential
breaking lepton number conservation is
\begin{equation}
W_{R_p \hspace{-0.8em}/\;\:} = \lambda_{ijk}L_i L_j {\bar E}_k +
\lambda'_{ijk}L_i Q_j {\bar D}_k.
\label{R-viol}
\end{equation}
Here $L$,  $Q$ are lepton and quark
doublets while  ${\bar E},\  {\bar D}$
are lepton and {\em down} quark singlet superfields.
Indices $i,j, k$ denote generations and
$\lambda_{ijk} = - \lambda_{jik}$.
In what follows we concentrate on the so called
"direct" SUSY contribution to the ${0\nu\beta\beta}$
\cite{Mohapatra}-\cite{HKK} depending only on
the $\lambda'$ term of the superpotential in Eq.\ (\ref{R-viol}).
The combination of both $\lambda'$ and $\lambda$ terms may lead to
the "indirect" SUSY contribution accompanied by
the neutrino exchange \cite{I_susy}, which we do not consider
in the present letter.

\begin{figure}[b]
\vspace{0.5cm}
\hspace{1.5cm}
\mbox{\epsfxsize=8.0 cm\epsffile{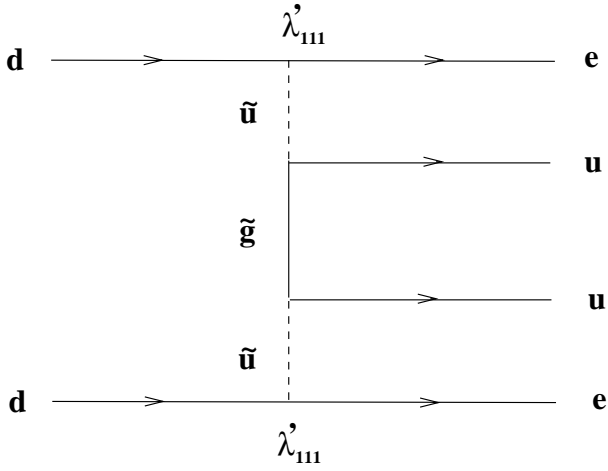}}
\vspace{0.5cm}
\caption{
An example of the supersymmetric contribution
to ${0\nu\beta\beta}$-decay.}
\label{fig2}
\end{figure}

Starting from the $\lambda'$ term in Eq.\ (\ref{R-viol})
the following  effective quark-electron vertex
has been derived \cite{HKK}:
\begin{eqnarray}
 {\cal L}_{qe}\ &=&
\frac{G_F^2}{2 m_{_p}}~ \bar e (1 + \gamma_5) e^{\bf c}
\nonumber \\
&\times & \left[(\eta_{\tilde q} + \eta_{\tilde f})
(J_{P}J_{P} + J_{S}J_{S})
- \frac{1}{4} \eta_{\tilde q}  J_T^{\mu\nu} J_{T \mu\nu}) \right].
\label{ql}
\end{eqnarray}
These interactions violate the electron number $\Delta L_e = 2$.
They are induced by  heavy SUSY particles in a virtual
intermediate state.
An example of the Feynman diagram contributing to
${\cal L}_{qe}$ is given in Fig.\ \ref{fig2}.
%
The color-singlet hadronic currents in Eq.\ (\ref{ql}) are
$
J_{P} =   \bar u^{\alpha} \gamma_5 d_{\alpha}, \
J_{S} =   \bar u^{\alpha} d_{\alpha}, \
J_T^{\mu \nu} = \bar u^{\alpha} \sigma^{\mu \nu}(1 + \gamma_5) d_{\alpha}$,
where $\alpha$ is a color index.

The lepton number violating parameters $\eta$ in Eq.\ (\ref{ql})
can be written in the following form
\begin{eqnarray}
\eta_{\tilde q} & = &
\Lambda^2\left(2 \alpha_s\frac{m_{_p}}{m_{\tilde g}} +
\frac{3}{4}\alpha_2 \frac{m_{_p}}{m_{\chi}}
(\epsilon_{Rd}^2 +\epsilon_{Lu}^2)\right),
\label{eta0}\\
\eta_{\tilde f} & = &
\Lambda^2\left(2 \alpha_s\frac{m_{_p}}{m_{\tilde g}} +
\frac{3}{2}\alpha_2 \frac{m_{_p}}{m_{\chi}}
\left(\frac{m_{\tilde q}}{m_{\tilde e}}\right)^2
{\cal C}\right).
\label{eta}
\end{eqnarray}
Here $ \Lambda =
(\sqrt{2\pi}/3) \lambda'_{111} G_F^{-1} m_{\tilde q}^{-2}$ and
$
{\cal C} = 6 (m_{\tilde q}/m_{\tilde e})^2 \epsilon_{Le}^2 -
\epsilon_{Rd}\epsilon_{Le} -
\epsilon_{Lu}\epsilon_{Rd}(m_{\tilde e}/m_{\tilde q})^2
- \epsilon_{Lu}\epsilon_{Le}
$.
$\alpha_2 = g_{2}^{2}/(4\pi)$ and $\alpha_s = g_{3}^{2}/(4\pi)$ are
$SU(2)_L$ and $SU(3)_c$ gauge coupling constants;
$m_{\tilde g}$ and $m_{\chi}$ are masses
of the gluino $\tilde g$ and of the lightest neutralino $\chi$.
The latter is a linear combination of the gaugino and higgsino
fields
$\chi = \alpha_{\chi} \tilde{B} +  \beta_{\chi} \tilde{W}^{3} +
\delta_{\chi} \tilde{H}_{1}^{0} + \gamma_{\chi} \tilde{H}_{2}^{0}$.
Here $\tilde{W}^{3}$ and $\tilde{B}$ are neutral
$SU(2)_L$ and $U(1)_Y$ gauginos while $\tilde{H}_{2}^{0}$,
$\tilde{H}_{1}^{0}$
are higgsinos which are the superpartners of the two neutral Higgs boson
fields $H_1^0$ and $H_2^0$
with a weak hypercharge $Y=-1, \ +1$, respectively.
The mixing coefficients $\alpha_{\chi},\beta_{\chi},\gamma_{\chi},
\delta_{\chi}$ can be obtained from diagonalization of the $4\times 4$
neutralino mass matrix.
Neutralino couplings are defined as 
$\epsilon_{L\psi} = - T_3(\psi) \beta_{\chi} +
\tan \theta_W \left(T_3(\psi) -  Q(\psi)\right) \alpha_{\chi}$,
$\epsilon_{R\psi} = Q(\psi) \tan \theta_W \alpha_{\chi}$  \cite{Haber}.
Here $Q $ and $ \ T_3$ are the electric charge and weak
isospin of the fields $\psi = u, d, e$.
In Eqs.\ (\ref{eta0})-(\ref{eta}) we used the universal squark
mass $m_{\tilde q}$  ansatz at the weak scale
$m_{\tilde u}\approx m_{\tilde d}\approx m_{\tilde q}$.
This approximation is justified by the constraints from
the flavor changing neutral currents and is sufficient for our analysis.

Now we have to reformulate the quark-lepton interactions in Eq.\ (\ref{ql})
in terms of the effective hadron-lepton interactions, which is necessary
for the further nuclear structure calculations.
The effective Lagrangian taking into account both
the nucleon (p, n) and $\pi$-meson degrees of freedoms in a nucleus can
be written as follows:\\[-3mm]
\begin{eqnarray}
  {\cal L}_{he} &=& {\cal L}_{2N} + {\cal L}_{\pi e} + {\cal L}_{\pi N}
\nonumber \\
&=& \frac{G_F^2}{2 m_{_p}}~
\bar p \Gamma^{(i)} n ~ \bar p \Gamma^{(i)}  n
~\bar{e} (1 + \gamma_5) e^c
\nonumber \\
&&- \frac{G_F^2}{2 m_{_p}}~ m_{\pi}^4  a_{\pi} \left(\pi^-\right)^2
~\bar{e} (1 + \gamma_5) e^c
\nonumber \\
&&+ g_{_s}~ \bar p\ i \gamma_5 n ~ \pi^+.
\label{2pi}
\end{eqnarray}
Here ${\cal L}_{2N}$,  ${\cal L}_{\pi e}$ and ${\cal L}_{\pi N}$
describe the conventional two-nucleon mode, pion-exchange mode and
pion-nucleon interactions, respectively.
They correspond to the first, the second
and third terms of the second part of the equation (\ref{2pi}).
$g_{_s} = 13.4\pm 1$ is known from experiment.
The two-nucleon mode contributions ${\cal L}_{2N}$
 to the ${0\nu\beta\beta}$-decay
with different operator structures $\Gamma^{(i)}$ were derived
and studied in  \cite{Vergados}-\cite{HKK}
within the $R_p \hspace{-1em}/\;\:$ MSSM.

In this note we concentrate on the effect of the pion-exchange term
${\cal L}_{\pi e}$.
The basic parameter $a_{\pi}$ of the Lagrangian ${\cal L}_{\pi e}$
can be approximately related to the parameters of the fundamental Lagrangian
${\cal L}_{qe}$ using the on-mass-shell "matching condition"
$
<\pi^+|{\cal L}_{qe}|\pi^-> =
<\pi^+|{\cal L}_{\pi e}|\pi^->
$.
The solution of this equation is
$a_{\pi} =
\frac{1}{2} (\eta_{\tilde q} + \eta_{\tilde f}) (c_{_P} + c_{_S})
- \frac{1}{8} \eta_{\tilde q} c_{_T}$,
where
$<\pi^+|J_i J_i|\pi^-> = - m_{\pi}^4 c_{i}$ with $i=P,S,T$.
Thus, we obtain the approximate hadronic "image"
${\cal L}_{\pi e}$ of the fundamental quark-lepton Lagrangian
${\cal L}_{qe}$ given in Eq.\ (\ref{ql}).

The contribution of the $J_{P,S,T}$ currents to $a_{\pi}$
can be estimated within the vacuum insertion approximation (VIA).
Applying Partial Conservation of Axial Current  (PCAC) we obtain
\begin{eqnarray}
<\pi^+|J_P J_P|\pi^-> &=&
\frac{8}{3}<\pi^+| J_{P}|0><0| J_{P} |\pi^-> \nonumber \\
&=&- \frac{16}{3} f_{\pi}^2 \frac{m_{\pi}^4}{(m_u + m_d)^2} \equiv
- m_{\pi}^4 c_{_P}.
\label{vacuum}
\end{eqnarray}
where $8/3$ is a combinatorial color factor and
$f_{\pi} = 0.668~ m_{\pi}$.
Taking the conventional values of the current quark masses
$m_u=4.2$ MeV, $m_d = 7.4$ MeV one gets $c_{_P}\approx 342$.
Within the VIA we have $c_{_S} = c_{_T} = 0$ since
$<0| J_{S}|\pi(p_{\pi})> = <0| J_{T}^{\mu\nu}|\pi(p_{\pi})> = 0$.
The scalar  current matrix element vanishes due to the parity arguments,
the tensor one vanishes due to
$ J_{T}^{\mu\nu} = -  J_{T}^{\nu\mu}$ and the impossibility of
constructing antisymmetric object having  only one
4-vector $p_{\pi}$. Thus, we expect the $J_P$ contribution to be
dominant.

The $J_P$ dominance also follows from  the non-relativistic quark model (QM)
\cite{Adler}.
Within this model one can calculate $<\pi^+|J_P J_P|\pi^->$  using
the closure approximation for the intermediate meson states ~\cite{Vergados1}.
After quite tedious calculations we end up again
with $c_{_P} >> c_{_{S,T}}$.
In this case the numerical value $c_{_P}\approx 1100$ is larger than
that in Eq.\ (\ref{vacuum}) since in addition to the vacuum
state there are other intermediate states taken into account.
In what follows we use both the VIA and the QM values
of $c_{_P}$.

The large coefficient $c_{_P}$ enhances
the pion-exchange contribution to the ${0\nu\beta\beta}$-decay.
This enhancement factor
is a generic property of the $R_p \hspace{-1em}/\;\:$
 SUSY-models  generating at low energies
the $J_P J_P$ interactions [see Eq.\ (\ref{ql})].
There is another factor enhancing the pion-exchange contribution
compared to the two-nucleon mode. As explained latter on,
it stems from the fact that the pion-exchange is longer ranged and
thus covers
a larger interval of the inter-nucleon distances enhancing
the nuclear matrix elements over the two-nucleon mode.

Starting from the Lagrangian ${\cal L}_{he}$ in Eq.\ (\ref{2pi})
it is straightforward to calculate
the contribution to the ${0\nu\beta\beta}$-matrix element
${\cal R}_{{0\nu\beta\beta}}$
which corresponds to the fundamental vertex ${\cal L}_{qe}$
in Eq.\ (\ref{ql}).
It consists of the two terms
${\cal R}_{{0\nu\beta\beta}} =
{\cal R}_{{0\nu\beta\beta}}^{2N} + {\cal R}_{{0\nu\beta\beta}}^{\pi N}$
describing the conventional two-nucleon mode
${\cal R}_{{0\nu\beta\beta}}^{2N}$ and
the pion-exchange contribution ${\cal R}_{{0\nu\beta\beta}}^{\pi N}$.
The relevant Feynman diagrams are given in the Fig.\ \ref{fig1}.
The corresponding half-life formula
reads
\begin{eqnarray}
\lefteqn{
\big[ T_{1/2}^{{0\nu\beta\beta}}(0^+ \rightarrow 0^+) \big]^{-1} =
 G_{01} \left(\frac{m_A}{m_{_p}}\right)^4 }\nonumber \\
&&\times \left | \eta_{\tilde q}~ {\cal M}_{\tilde q}^{2N}
 +  \eta_{\tilde f}~ {\cal M}_{\tilde f}^{2N} +
(\eta_{\tilde q} + \eta_{\tilde f})~ {\cal M}^{\pi N} \right |^2.
\label{half-life}
\end{eqnarray}
Here $G_{01}$ is the standard phase space factor tabulated
for various nuclei in  \cite{doi85} and $m_A = 850$ MeV.
The analytic form of the two-nucleon mode nuclear matrix elements
${\cal M}_{\tilde q, \tilde f}^{2N}$ are given in \cite{HKK}.
Here we present the new pion-exchange nuclear matrix element defined as
\begin{eqnarray}
{\cal M}^{\pi N} = \frac{m_{_p}}{ m_e}
      \alpha^{\pi}\left( {\cal M}_{GT,\pi} + {\cal M}_{T,\pi} \right),
\label{nme}
\end{eqnarray}
The partial Gamow-Teller and tensor matrix elements are
\begin{eqnarray}
{\cal M}_{GT,\pi} &=& <0^+_f| \sum^{}_{i\neq j}
{\tau}^{+}_{i}{\tau}^{+}_{j}\sigma_{ij}
\left(\frac{R}{r_{ij}}\right) F_{1}(x_{\pi})
|0^+_i>,\\
{\cal M}_{T,\pi} &=&
 <0^+_f| \sum^{}_{i\neq j}
{\tau}^{+}_{i}{\tau}^{+}_{j}S_{ij}
\left(\frac{R}{r_{ij}}\right) F_{2}(x_{\pi}) |0^+_i>,
\label{pion-part}
\end{eqnarray}
where
\begin{eqnarray}
\lefteqn{S_{ij}= 3\vec{\sigma}_i\cdot \hat{\vec{r}}_{ij}
         ~\vec{\sigma}_j\cdot \hat{\vec{r}}_{ij}
       -\vec{\sigma}_i\cdot \vec{\sigma}_j,~~
\sigma_{ij}=\vec{\sigma}_i\cdot \vec{\sigma}_j ,
} \nonumber \\
&&\hat{\vec{r}}_{ij} = (\vec{r}_i - \vec{r}_j)/|\vec{r}_i - \vec{r}_j|,~
r_{ij}=|\vec{r}_i - \vec{r}_j|
\end{eqnarray}
and $x_{\pi} = m_{\pi} r_{ij}$.
Here  $\vec{r}_i$ is the coordinate of the "i-th" nucleon.
The pion-nucleon structure coefficient
in Eq.\ (\ref{nme}) is given by
\begin{eqnarray}
\alpha^{\pi} = \frac{1}{96}
\left(\frac{m_A}{m_{_p}}\right)^2\left(\frac{m_{\pi}}{m_A}\right)^4
\left(\frac{g_s}{f_A}\right)^2  c_{_P},
\label{al_pi}
\end{eqnarray}
where $f_A = 1.261$.
The pion-exchange SUSY potentials are
\begin{eqnarray}
F_1(x) = (x - 2) e^{- x}, \ \ \ F_2(x) = (x + 1) e^{- x}.
\label{potentials}
\end{eqnarray}

The most stringent experimental lower limit
on the $0\nu\beta\beta$-decay  half-life has been 
obtained for $^{76}{\text{Ge}}$ \cite{hdmo94},
what favors especially this nucleus for nuclear structure
calculations. In this letter we pay our attention only to this
isotope. We have employed the renormalized quasiparticle random phase
approximation with proton-neutron pairing (full-RQRPA) \cite{FRQRPA}
to calculate both the two-nucleon and pion-exchange
nuclear matrix elements governing the $R_p \hspace{-1em}/\;\:$
 SUSY $0\nu\beta\beta$-decay of $^{76}{\text{Ge}}$.
The full-RQRPA includes the Pauli effect of fermion pairs and
does not collapse for a physical value of the nuclear force strength.
To include the Pauli principle more correctly we do not use
the quasi-boson-approximation to derive the Quasiparticle Random
Phase Approximation (QRPA). If one includes the exact Fermion
commutation relations for nucleon pairs (two quasiparticles)
as a QRPA expectation value, one obtains the renormalized QRPA
(RQRPA), which is stable against the collapse of $2\nu\beta\beta$
Gamow-Teller transition. Therefore the RQRPA offers a significantly
more reliable treatment of the nuclear many-body problem for the
description of the $0\nu\beta\beta$ decay.
Thus it also allows one to establish more reliable constraints on
the $R_p \hspace{-1em}/\;\:$
 SUSY parameters from the best available experimental
lower bound on the  $0\nu\beta\beta$-decay half-life.
We have found the following numerical
values of the nuclear matrix elements for $^{76}{\text{Ge}}$:
$
{\cal M}_{\tilde q}^{2N} = -61; \
{\cal M}_{\tilde f}^{2N} = 0.85$; \
${\cal M}^{\pi N}        = -1800$(QM), $-600$(VIA).
The pion-exchange matrix element is given for QM
(non-relativistic Quark Model)
and VIA (Vacuum Insertion Approximation) values
of the coefficient $c_{_P}$.
It is apparent that in both QM and VIA cases the dominant
contribution to Eq.\ (\ref{half-life})
comes from the pion-exchange mechanism corresponding to ${\cal M}^{\pi N}$.
The VIA value ${\cal M}^{\pi N} = -600$
we will use for conservative estimations.
It is worthwhile to note that the above nuclear matrix elements
are quite stable with respect to variation of the nuclear model parameters.
The uncertainty of the calculated values of
${\cal M}_{\tilde q, \tilde f}^{2N}$
and ${\cal M}^{\pi N}$ does not exceed 20\%.

Now we are ready to extract the constraints on the
$R_p \hspace{-1em}/\;\:$ MSSM parameters
from the non-observation of ${0\nu\beta\beta}$-decay.
The current experimental lower bound on the $^{76}{\text{Ge}}$
${0\nu\beta\beta}$-decay
half-life \cite{hdmo94} is
$T_{1/2}^{{0\nu\beta\beta}-\text{exp}}(0^+ \rightarrow 0^+)
\hskip2mm \geq \hskip2mm
9.1 \times 10^{24}$ years $90 \% \ \text{c.l.}
$
Combining this bound with Eq.\ (\ref{half-life}) and the above given
numerical values of ${\cal M}^{\pi N}$ we get a
constraint on the sum of the effective MSSM parameters
$\eta_{\tilde q} + \eta_{\tilde f} \leq 2.1\times 10^{-9}$. If one
does not include the pion-exchange contribution then
one gets a constraint
$\eta_{\tilde q} + 0.014\eta_{\tilde f} \leq 7.8\times 10^{-8}$
from the remaining 2N-mode.
It is essentially less stringent than the above given
$\pi$N-mode constraint by more than one order of magnitude
for  $\eta_{\tilde q}$  and by three orders for $\eta_{\tilde f}$.

The gluino and neutralino contributions to $\eta_i$ can not cancel
each other within the present experimental limits on their masses
and couplings [see  \cite{HKK}].
Therefore, we can extract from the above limit on $\eta_i$
the constraints on these individual contributions.
The gluino contribution constraint is
\begin{equation}
\lambda'_{111} \le 2.0(1.18)\times 10^{-4}
\Big({m_{\tilde q}\over{100 ~{\text{GeV}}}} \Big)^2
 \Big({m_{\tilde g}\over{100 ~{\text{GeV}}}} \Big)^{1/2},
\label{constr1}
\end{equation}
for the VIA(QM) value of the pion matrix element parameter $c_{_P}$.
The neutralino contribution constraint is more complex because
it involves more parameters: neutralino mixing coefficients, selectron and
squark masses. However, it can be cast into the form of Eq.\ (\ref{constr1})
under the phenomenologically viable simplifying assumptions.
Assume that the neutralino is B-ino dominant
$\alpha_{\chi}>> \beta_{\chi},\delta_{\chi},\gamma_{\chi}$ and that
$m_{\tilde q}\geq m_{\tilde e}$.
Then we get
\begin{equation}
\lambda'_{111} \le 5.2(3.07)\times 10^{-4}
\Big({m_{\tilde e}\over{100 ~{\text{GeV}}}} \Big)^2
 \Big({m_{\chi}\over{100 ~{\text{GeV}}}} \Big)^{1/2}.
\label{constr2}
\end{equation}
If all SUSY particle masses in
Eqs.\ (\ref{constr1})-(\ref{constr2}) were at their present experimental
lower bounds ~\cite{RPP} $m_{\tilde q} \geq 90$ GeV,
$m_{\tilde e}\geq 45$ GeV,
$m_{\tilde g}\geq 100$ GeV, $m_{\chi}\geq 19$ GeV, we could estimate
the size of the $R_p \hspace{-1em}/\;\:$ coupling constant
$
\lambda'_{111} \le 4.6(2.7)\times 10^{-5}
$.
A conservative bound can be obtained using the SUSY "naturalness"
upper bound $m_{\tilde g,\tilde q,\chi}\leq 1$TeV.
It gives
$
\lambda'_{111} \le 6.3(3.7)\times 10^{-2}
$.
This limit and those in Eqs.\ (\ref{constr1})-(\ref{constr2})
are the best known
limits on the $R_p \hspace{-1em}/\;\:$ coupling $\lambda'_{111}$
[see  \cite{HKK} and references therein].

It is interesting to compare the ${0\nu\beta\beta}$-decay
and accelerator experiments
from the point of view of their sensitivity to the
$R_p \hspace{-1em}/\;\:$ SUSY signal.
Previously  \cite{HKK} it was shown that even in the 2N-mode
the constraints from the
 ${0\nu\beta\beta}$-decay exclude the domain of
the $R_p \hspace{-1em}/\;\:$  MSSM parameter space accessible for the ongoing experiments
with the ZEUS detector at  HERA. This conclusion touched upon the region
of the so called $R_p \hspace{-1em}/\;\:$
resonant single squark $\tilde q$ production
mechanism in deep inelastic ep-scattering ~\cite{drein1}.
Taking into account the $\pi$N-mode makes this conclusion
much stronger.
The reason is that the excluded region becomes significantly larger than
in case of the 2N-mode alone. Now this region extends so far that
it would include the HERA domain even if the experimental
lower bound on $T_{1/2}^{0\nu\beta\beta}$ was
by about (conservatively) five orders of magnitude less than the currently
existing one.

Nevertheless, there is
still a window for the HERA experiments in the region corresponding to
another mechanism assuming $R_p$-conserving $\tilde e + \tilde q$ production
and their subsequent $R_p \hspace{-1em}/\;\:$
cascade decays. This region is
$m_{\tilde e} + m_{\tilde q}\leq 205$ GeV and $\lambda'_{111}\geq 10^{-6}$
~\cite{drein2}\footnote{Note that unlike
${0\nu\beta\beta}$-decay searches HERA
can probe within this mechanism $R_p \hspace{-1em}/\;\:$
 coupling constants other than $\lambda'_{111}$.}.
In the present letter we would like to stress that
the ${0\nu\beta\beta}$-decay constraints
given in Eqs.\ (\ref{constr1})-(\ref{constr2})
dramatically reduces the above quoted region.
Only a narrow part of it corresponding
to very low values of
the $R_p \hspace{-1em}/\;\:$ coupling constant
$10^{-6}\leq \lambda'_{111}\leq 0.76(2.4)
\cdot 10^{-4} (\frac{m_{\tilde e} + m_{\tilde q}}{100 ~GeV})^2$
and $m_{\tilde e} + m_{\tilde q}\leq 205$ GeV
remains not excluded by the ${0\nu\beta\beta}$-decay constraints.
Here we used
the conservative VIA value of the pion matrix element $c_P$ and put
$m_{\chi} = m_{\tilde g} =100$ GeV (1 TeV).

Summarizing, we point out that the SUSY contribution
to the $0\nu\beta\beta$-decay comes dominantly via
the pion-exchange mechanism considered in the present letter.
The conventional two-nucleon mechanism
\cite{Mohapatra}-\cite{HKK}, corresponding to
$nn\rightarrow ppee$ transition without
light particles (pion or neutrino) in the intermediate state,
brings only a subdominant SUSY contribution.
On practice the pion-exchange mechanism considerably
enhances the sensitivity of the ${0\nu\beta\beta}$-decay to
the supersymmetry.
This allowed us to obtain presently the most stringent limitations of
the certain first generation $R_p \hspace{-1em}/\;\:$  MSSM parameters.

\acknowledgements
We thank V.A. Bednyakov, V.B. Brudanin,
M. Hirsch and H.V. Klapdor-Kleingrothaus
for fruitful discussions.
F.S. and S.K. would like to thank
the "Deutsche Forschungsgemeinschaft" for financial
support by grants Fa 67/17-1 and 436RUS17/137/95, respectively.

\newpage

\bigskip

\bigskip

\bigskip

\bigskip

\end{document}